\documentstyle[11pt,newpasp,twoside,epsf]{article}
\def\edcomment#1{\iffalse\marginpar{\raggedright\sl#1\/}\else\relax\fi}
\reversemarginpar

\begin{document}
\title{The Energy of Gamma-Ray Bursts}
 \author{Tsvi Piran}
\affil{Racah Institute of Physics, The Hebrew University,
Jerusalem, Israel}

\begin{abstract}
The energy of a Gamma-Ray Burst is one of the most interesting
factors that can help determining the origin of these mysterious
explosions.  After the discovery that GRBs are cosmological it
was thought, for a while that they are standard candles releasing
$\sim 10^{51}$ergs. Redshift measurements that followed the
discovery of GRB afterglow revealed that  GRBs  are (i) further
than what was initially believed and (ii) have a very wide
luminosity function. Some bursts revealed an enormous energy
release with a record estimate of $ 1.4 \times 10^{54}$ergs for
GRB990123. The energy budget was stretched even further when it
was suggested that the conversion efficiency of producing
gamma-rays is low.  The realization that GRBs are beamed changed
this perspective and brought the energy budget of GRBs down back
to a "modest" $\sim 5 \times 10^{50}$ergs. I discuss here various
estimates showing that  GRB energy is narrowly distributed  and
discuss the implications of this conclusion to GRB models.
\end{abstract}

\section{Introduction}

The  energy release in a Gamma ray burst is one of the best clues
on the nature of these objects. However, this simple and basic
quantity is rather hard to get. The observed flux was readily
available even for the first bursts. But the lack of a reliable
distance estimate was a series obstacle. When redshift
measurments became available it turned out that even this
knowledge was not sufficient. Other complications arose. On one
hand theoretical considerations showed that the efficiency of
conversion of the energy to $\gamma$-rays could not be very high
and estimates based on $\gamma$-rays alone were too low. On the
other hand, it turned out that GRBs are beamed and estimates that
assumed isotropic  emission were too high. In this lecture I
discuss the current understanding of the energetics of GRBs and
the implication to models of the "inner engines" that power GRBs.

In the first GRB revolution BATSE (Meegan et al., 1992)
demonstrated that GRBs are cosmological. This has changed the
distance scale by six orders of magnitude and the energy scale by
twelve orders of magnitude. BATSE's count distribution for long
bursts ($T_{90}>2$sec)  is consistent with a cosmological standard
candle distribution (Cohen \& Piran, 1995) with $E\approx
10^{51}$ergs. This has led to the believe that (at least long)
GRBs are standard candles and that the bursts are observed by
BATSE up to $z\approx 2$.

The second GRB revolution took place in 1997 with the discovery of
GRB afterglow (Costa et al, 1997; van Paradijs et al., 1997). The
Italian-Dutch satellite BeppoSAX provided angular position of
several dozen long bursts to within about 3 arc-minutes which
enabled follow up observations  in the x-ray (see Piro, 2000),
optical, milli-meter and radio frequencies which has provided a
wealth of information on these explosions. In several cases the
redshift of the afterglow or the host galaxy could be measured.
This provided a new and direct estimate of the distance and of
the energy involved. The first redshift, $z=0.835$ was obtained
for GRB970508. The corresponding energy, $5.4 \times 10^{51}$ergs,
was more or less in line with the standard candles estimates!

It quickly became apparent that GRBs are not standard candles.
With a  redshift 3.4 and energy $2 \times 10^{53}$ergs GRB971214
was too far and too energetic for the standard candle picture.
GRB990123 was even more energetic. It turned out that the GRB
energy distribution is very broad and that their rate follows the
star formation rate ( Totani, 1997; Sahu et al, 1997; Wijers et
al, 1998). The consistency of a standard candle cosmological
model with BATSE's peak counts distribution remains an
inexplicable coincidence.

The observations of extremely large energies (at least in some
bursts) suggested that some GRBs involve energy release of
stellar mass or more, ruling out several of the leading models at
the time. The situation was even more worrisome when one
considered the issue of efficiency.  Most models and in particular
the  internal shocks model, cannot produce $\gamma$-rays at 100\%
efficiency. There was a concern that the efficiency of the
internal shocks model is rather low (at the level of a few
percent (Kobayashi et al, 1997; Daigne, \& Mochkovitch, 1998).
This increased further the energy requirement for GRBs, ruling
out most models and leading to a GRB energy crisis (Kumar, 1999).

On the other hand GRBs could be beamed. The energy budget would
then drop down significantly, by a factor $\theta^2 /2$, where
$\theta$ is the opening angle of the jet. Rhoads (1999) pointed
out that an expanding relativistic jet would exhibit spherical
like evolution as long as $\Gamma > \theta^{-1}$. It will expand
rapidly sideways afterwards. This behaviour will produce a break
in the afterglow light curve. Using this break one could determine
the opening angles and estimate the true energy budget. Already in
1999 Sari, Piran \& Halpern pointed out that the two most
energetic GRBs (at that time) exhibit jet-like behaviour. This
suggested that beaming is a strong factor in the energetics of
GRB.

I summarize here several different estimates to the energy of
GRBs. I argue that there are good indications that different GRBs
emit relatively constant energy. GRBs are standard candles after
all. However, differences in beaming factors lead to a variation
of three orders of magnitude in the isotropic equivalent energies
and to a very wide apparent luminosity function.

\section{The Overall Picture and the Fireball Model}

Following the discovery of GRB afterglow  we have learned during
the last four  years a great deal about long duration gamma-ray
bursts (GRBs).   These observations are described well by the
relativistic fireball model (see e.g. Piran, 1999). According to
this model the energy from the central source is deposited in
material that moves  very close to the speed of light. The
kinetic energy of this material is converted to the observed
$\gamma$-rays as a result of collisions between fast moving
material that catches up with slower moving ejecta. The afterglow
is produced latter by the shock heated circum-burst medium (see
Fig. 1). The nature of the ``inner engines" that expels
relativistic material, which is responsible for the the GRBs, is
not determined yet. I discuss here the energetic of GRB sources in
the context of this model. One of the problems in discussing the
energy of GRBs is that there are several different energies
involved. I begin with a discussion of the different energy
definitions..

\begin{figure}
\plotfiddle{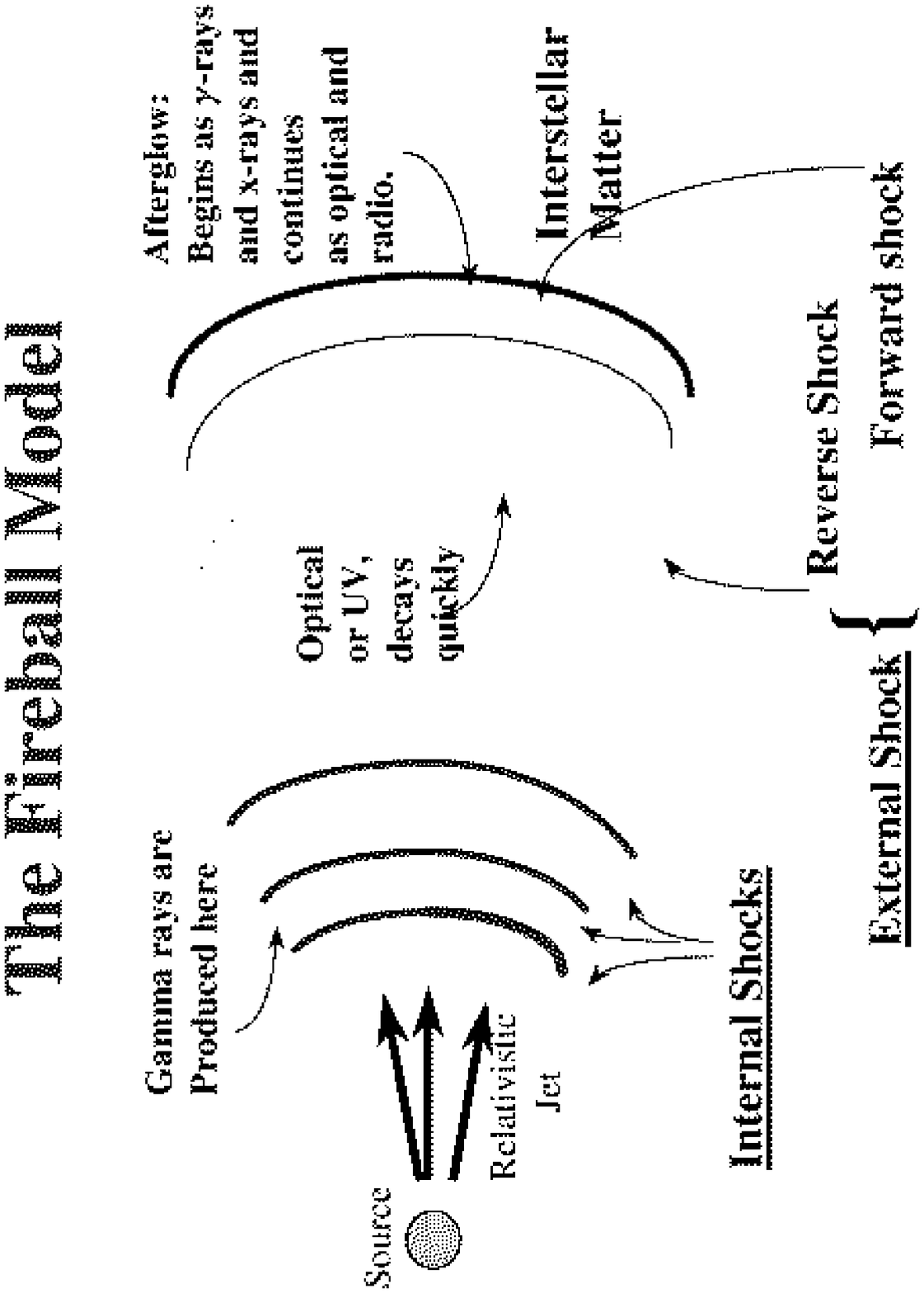}{2.6truein}{-90}{50}{45}{-190}{250}
\caption{\footnotesize{The internal-external
shocks fireball model: The GRB is produced by internal shocks
within the flow. The afterglow is produced by shock heated
circum-burst matter.}}
\end{figure}

\subsection{Eneregy - Theoretical Considerations}

Our goal is to find the best estimate for
\begin{itemize}
\item {\bf $E_{tot}$}:
the total energy emitted by the source.
\end{itemize}
However, it is practically impossible to determine this energy.
It is not known what are all forms of energy release by the
source. For example, the fireball model is based the conversion
of the kinetic energy of relativistic flow (particles or Poynting
flux) to  $\gamma$-rays. It is impossible to detect
non-relativistic particles that may also be emitted by the
source. A second, more accessible quantity is
\begin{itemize}
\item {\bf $E_{rel}$}:
the energy of the relativistic flow emitted by the source.
\end{itemize}
A fraction, $\epsilon$, of $E_{rel}$ is converted via the
internal shocks to the observed prompt $\gamma$-ray emission. The
rest,  $(1-\epsilon)E_{rel}$,  is dissipated later via external
shocks on the circum-burst matter producing the afterglow. An
unknown fraction of this energy is dissipated  in the radiative
phase (which  has not been observed yet) during the first half
hour of the afterglow.  The remaining energy is:
\begin{itemize}
\item {\bf $E_{K}$}:
the kinetic energy during the adiabatic afterglow phase.
\end{itemize}
$E_{K}$ is dissipated gradually over a period of months or even
years producing  the observed afterglow. Clearly, we have:
$E_{K} < (1-\epsilon) E_{rel} \le E_{tot} \ $.
Observations of long time tails of GRBs suggest that  this losses
are small (Burenin et al, 1999, Giblin et al, 1999; Tkachenko et
al., 2000). Thus, $E_{K} \approx (1-\epsilon) E_{rel}$.  The
efficiency, $\epsilon$, cannot be too large otherwise there won't
be afterglow. It cannot be too small either, otherwise we will
reach a GRB energy crisis. Combining these facts we expect,
therefore, that $E_K \approx E_{rel}$ to within say a factor of a
few.

\subsection{Eneregy - Observational Considerations}
Given the observed $\gamma$-ray fluence and the redshift one can
easily estimate
\begin{itemize}
\item
$E_{\gamma,iso}$: the energy emitted in $\gamma$-rays  assuming
that the emission is isotropic.
\end{itemize}
$E_{\gamma,iso}$ can also be estimated from the BATSE catalogue
by fitting the flux distribution to theoretical models (Cohen \&
Piran, 1995; Schmidt, 2001). As afterglow observations proceeded,
alarmingly large values (Kulkarni et al. 1999) ($3.4 \times
10^{54}$ergs for GRB990123) were measured for $E_{\gamma,iso}$.
As we discuss shortly (Rhoads, 1999; Sari Piran \& Halpern, 1999)
GRBs are beamed and $E_{\gamma,iso}$ is far larger than the actual
energy emitted in $\gamma$-rays. We define instead:
\begin{itemize}
\item
$E_\gamma \equiv  (\theta^2/2)E_{\gamma,iso}$.
\end{itemize}
Here $\theta$ is the effective angle of $\gamma$-ray emission.

One would expect that $E_\gamma$ is a good estimate to $E_{rel}$
as: $E_\gamma = \epsilon E_{rel}$. However there are several
prooblems. First $\epsilon$ is unknown. Moreover one would expect
it to  vary significantly from one burst to another. Second, the
large Lorentz factor during the $\gamma$-ray emission phase,
makes the observed $E_\gamma$ rather sensitive to angular
inhomogeneities of the relativistic ejecta (Kumar \& Piran,
2000). During the GRB phase the relativistic Lorentz factor is at
least a few hundred (See e.g. Lithwick,\& Sari,  2001 for a
summary of the arguments concerning pair production opacity).
Thus, the observed $\gamma$-rays come from a region whose angular
size is $\gamma^{-1} \le 10^{-2}$. This is narrower by a factor
of ten than the angular width of the narrowest observed jets.
Thus the observed $\gamma$-rays span only a small fraction of the
actual emitting region. The estimated energy based on this data
alone could be misleading.

\subsection{The Efficiency of Internal Shocks}
The conversion efficiency of kinetic energy to $\gamma$-rays,
$\epsilon$, depends on two factors: the conversion efficiency of
bulk kinetic energy to energy of accelerated electrons and the
efficiency of radiating this energy to $\gamma$-rays. It is
usually assumed that this second factor is close to unity. The
first factor depends on the strength of the relevant shocks and
in turn on the relative Lorentz factors between the different
shells (Kobayashi, Piran \& Sari, 1997). There have been numerous
attempts to estimate this efficiency (Kobayashi, Piran \& Sari,
1997; Daigne, \& Mochkovitch, 1998; Kumar, 99; Spada et al, 2000;
Beloborodov, 2000; Kobayashi, \& Sari, 2001; Guetta et al, 2001)
and the results range from a few percent to near unity. The
conversion can be efficient (close to unity) if the distribution
of Lorentz factors is very wide (Beloborodov, 2000, Kobayashi, \&
Sari, 2001).  In those cases in which GRB afterglow is observed
this efficiency could not be too large. Otherwise there would be
no energy left to produce the afterglow.

\subsection{Jets and Beaming}

The original fireball model assumed spherical symmetry. As always
this was partially because of simplicity. However,  the spherical
approximation is valid as long as $ \Gamma > \Delta \theta^{-1}$,
where $\Delta \theta$ is the scale of angular inhomegeneity.
Because of time dilation sideways propagation is too slow and the
information on the inhomogeneity could not propagate (Piran,
1995). As the blast wave is slowed down by the circum burst
matter the Lorentz factor decreases and when $\Gamma \sim
theta^{-1}$ it begins to expand sideways (Rhoads, 1999). This
produces a break at the light curve, which is accompanied by a
change in the spectral index (Sari Piran \& Halpern, 1999).
Various models (usually analytic or semi-analytic) have been used
to describe the hydrodynamic evolution of this expanding phase
(Rhoads 1999; Sari Piran \& Halpern, 1999; Panaitescu \&
M\'esz\'aros 1999; Moderski, Sikora \& Bulik 2000; Kumar \&
Panaitescu 2000). Different assumptions have lead to different
relations between the opening angle and the $t_b$, the time of
the break in the light curve. Following Sari Piran \& Halpern
(1999) I adopt here :
\begin{equation}
\theta =0.12 (n/E_{51})^{1/8} t_{b,days}^{3/8}=0.052
(n/E_{K,51})^{1/6} t_{b,days}^{1/6} , \label{tb}
\end{equation}
where $E_{K,51}$ is the adiabatic kinetic energy in units of
$10^{51}$ergs, $E_{51}\equiv 2 E_{K,51}/\theta^2$ is the
isotropic-equivalent kinetic energy  and $t_{b,days}$ is the
break time in the afterglow light curve expressed in  days.

Detailed numerical simulation (Granot et al, 2001) show that the
sideway expansion is more complicated than what was assumed in
the simple analytic models (Fig. 2 depicts the density and the
velocity field from a jet long after the jet break. The sideway
expansion is not as prominent as expected.) Still, somewhat
surprisingly,  a jet break arises more or less at $t_b$ according
to Eq.1.

\begin{figure}
\plottwo{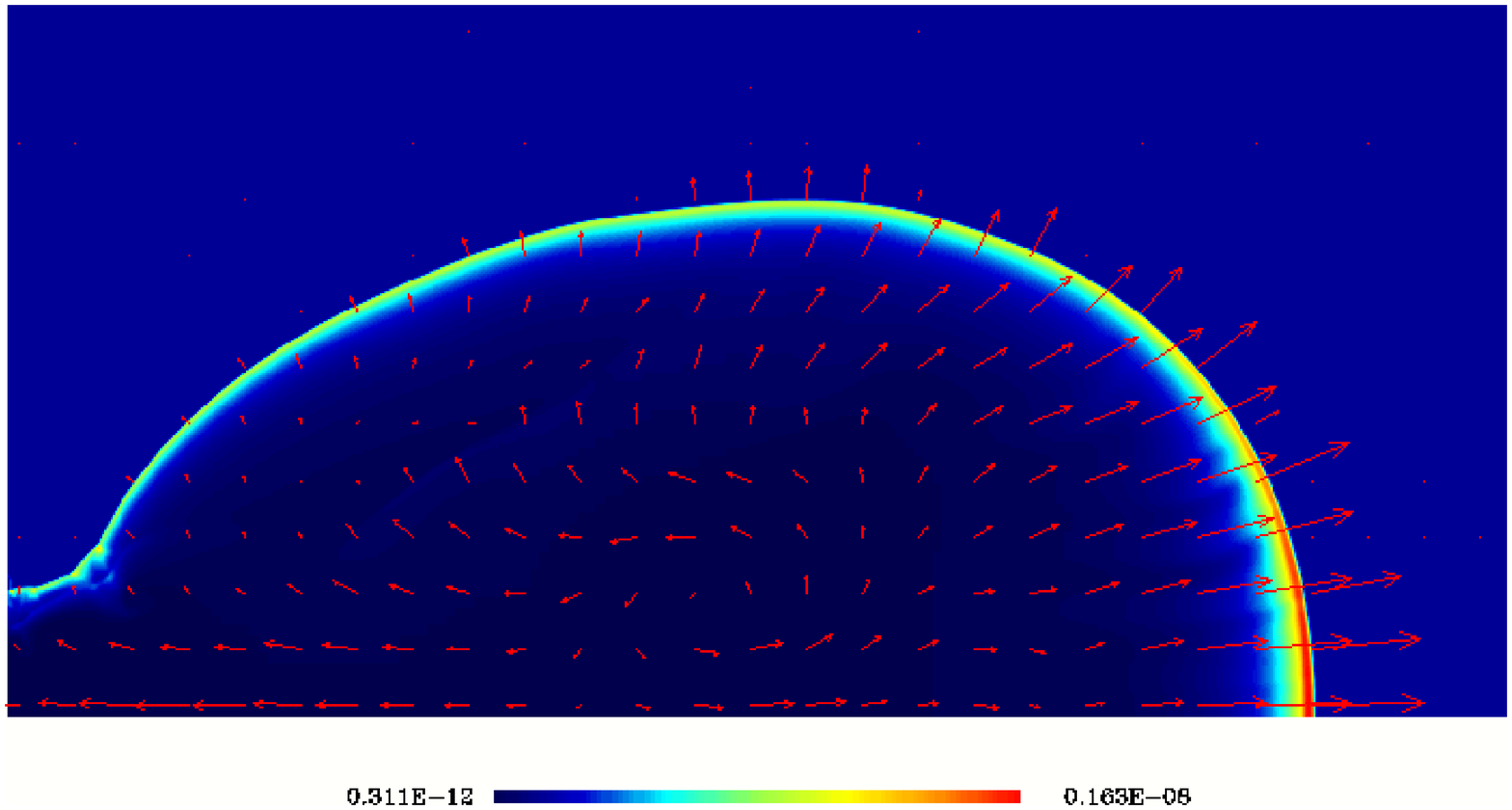}{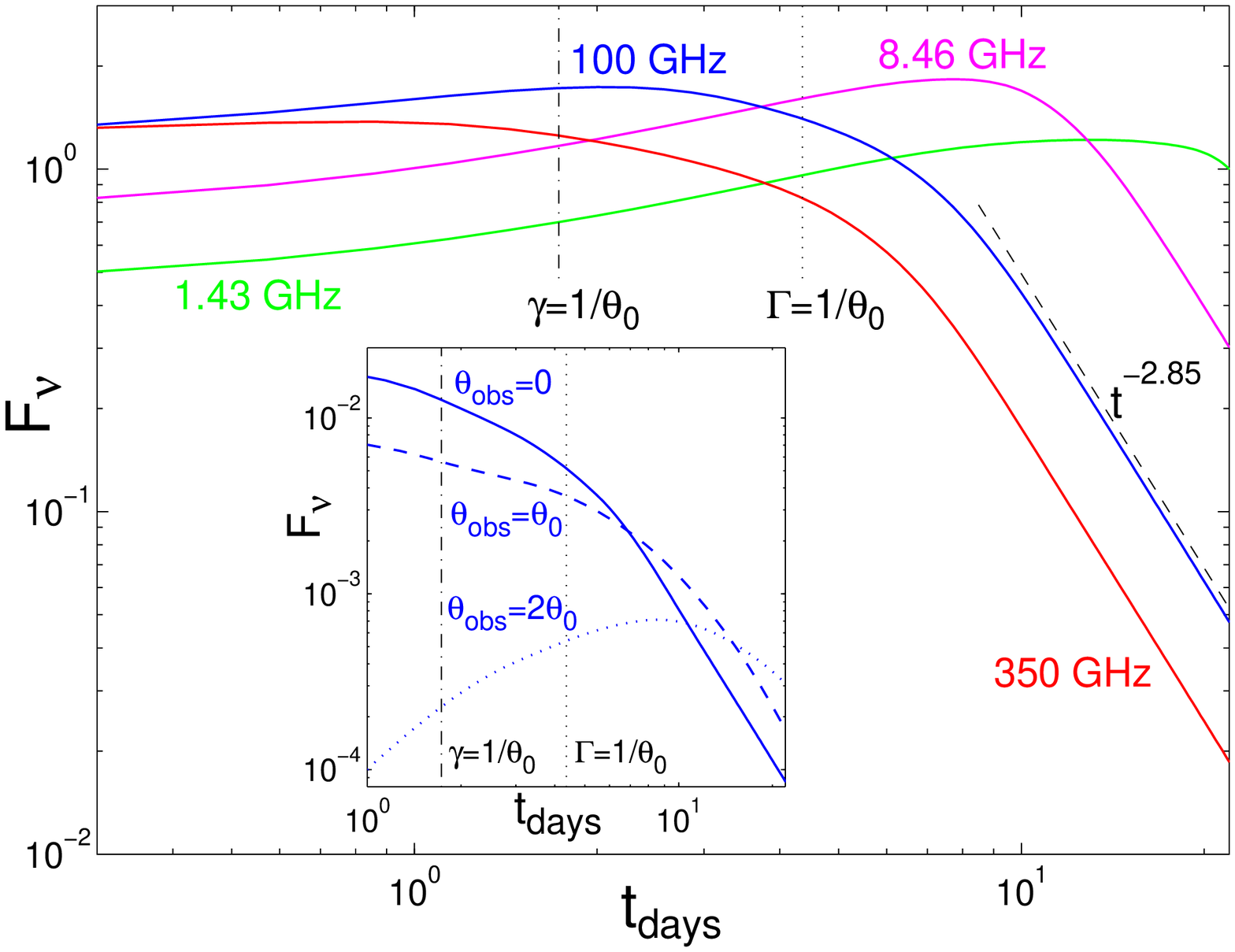} \caption{\footnotesize{From Granot
et al, (2001) {\bf Left:} Density countor and velocity fields in
an expanding jet long after the jet break.  {\bf Right}: Light
curves from a full hydrodynamic calculation of an expanding jet.
A clear break is seen, approximately at the time given by Eq. 1.}}
\end{figure}

\section{Energy Estimats}

The simplest and most direct estimates are of $E_{\gamma,iso}$.
Schmidt (1999) have fitted  the $ \langle V/V){max} \rangle $
distribution of BATSE's GRBs a cosmological model that follows
the star formation rate. He finds a break  energy of $1.2 \times
10^{53}$ergs in the 10-1000 keV band but as the higher slop is
steeper the average energy is $2.4 \times 10^{52}$ergs. This is of
course the isotropic energy. The energy estimate depends on the
specific cosmological model (Schmidt mentions variation by a
factor of 2.5 between different models), on the star formation
rate  and on the assumptions on the average spectral indices of
the GRB.

For GRBs with known redshifts one can estimate $E_{\gamma,iso}$.
Here care should be taken to obtain an exact estimate of the
spectrum (see  Jimenez  et al., 2001) and to add a proper K
correction (Bloom et al., 2001). Fig 3 depicts the isotropic
energy and the redshift for 17 bursts with known afterglow from
Bloom et al., (2001). One sees a large spread (as seen also in
Schmidt's (1999) luminosity function). Jimmenz et al (2001)
consider 8 BATSE bursts with a spectroscopic redshift estimate
and four additional bursts whose redshift has been estimated on
the basis of the host galaxy R magnitude. They find an average
$E_{\gamma,iso}$ of $1.3 \times 10^{53}$ergs. This agrees with
the average of Bloom's sample but is higher by a factor of 5 than
Schmidt's average for the full BATSE sample.

The next step would be to correct $E_{\gamma,iso}$ for beaming.
Frail et al. (2001, hereafter F1) estimated $E_\gamma$ for 18
bursts with redshift. They find a very narrow distribution with
typical values around $5 \times 10^{50}$ergs (see Fig 3) and FWHM
of a factor of 5. This energy estimate should be taken with care
as it depends critically on the estimated jet opening angles
$\theta_i$. Frail et al (2001, hereafter F01) estimate the
opening angles using Eq. 1. Panaitescu \& Kumar (2001, hearafter
PK01) perform a  detailed modeling of the afterglow and obtain
different estimates for $\theta_i$. A comparison of the opening
angles obtained by F01 and by PK01  for the same bursts shows
that while many of the estimates agree in two cases out of 6  the
angles differ by a factor of $\sim 3$. Furthermore one has to
worry about the possible angular inhomogeneity discussed earlier.
Given these facts the narrowness of the $E_\gamma$ distribution
is remarkable!

PK01 have  modeled the afterglow emission over a wide range of
frequency and time for 8 well studied bursts. The model is fitted
to the data and the fit yields several burst parameters,
including the adiabatic energy, $E_K$, and the jet opening angle.
The results are also shown on Fig 3. Using their estimate of the
opening angles $\theta_i$ PK01 also estimate $E_\gamma$ (see Fig.
3). The PK01 estimates for $E_\gamma$, differ from the F01
estimates for the same bursts by up to a factor of 8!

\begin{figure}
\plotfiddle{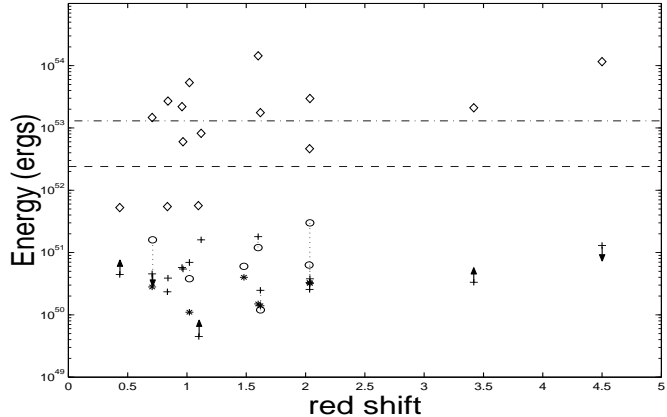}{2truein}{0}{50}{40}{-150}{-80}
 \caption{\footnotesize{$E_{\gamma,iso}$ (diamonds) from Bloom,
et al., (2001); $E_\gamma$ (crosses)  from F01,  and $E_\gamma$
(circles) and $E_{K}$ (stars) from PK01. $E_\gamma(F01)$  and
$E_\gamma(PK01)$ for the same burst are connected with a vertical
dotted line. The $E_{iso}$ distribution is consistent with the
average isotropic energy determined by Jimenez et al (2001) for
bursts with afterglows  (dashed-dotted line). Schmidts' (1999)
average $E_{\gamma,iso}$ is shown as a dashed line. Most
$E_{\gamma,iso}$ estimates for burst with  afterglow observations
are higher than Schmidts' average $E_{\gamma,iso}$}}
\end{figure}

\section{The Width of the Energy Distribution}

Both the F01 and PK01 results show that the energy distributions
are rather narrow. F01 find that the FWHM  of the $E_\gamma$
distribution is only a factor of 5. PK01 $E_{K}$ estimates are
very narrow, the width is only a factor of 3. The $E_\gamma$
estimates of PK01 are wider, but still narrow with a width of 11.
These values should be compared with the $E_{iso}$ distribution
that spans a factor of $10^4$. This motivates us to search for
different limits on the width of the energy distribution.

Following Piran et al (2001a) I turn now  to determine the spread
of $E_K$ using the x-ray afterglow flux.  The key factor in this
method is the observation that the x-ray luminosity at a fixed
time after the burst depends almost linearly on the kinetic
energy of the flow at that time and very weakly if at all on other
parameters (Kumar 2000; Freedman \& Waxman, 2001).

Consider a fireball with $\Gamma> \theta^{-1}$, so that the
spherical approximation holds. The standard synchrotron fireball
model implies that the isotropic equivalent luminosity at
frequency $\nu$ above the cooling frequency, at a fixed elapsed
time since the explosion, is given by (Kumar 2000; Freedman \&
Waxman, 2001):
\begin{equation}
L_x = \eta_p \left[{d E_K\over d\Omega}\right]^{(p+2)/4}
   \epsilon_e^{p-1}\epsilon_B^{(p-2)/4},
   \label{LX}
\end{equation}
where $dE_K/d\Omega$ is the kinetic energy per unit solid angle,
and $\eta_p$ is a constant. $L_x$  depends on the energy per unit
solid angle in the explosion, and on the fractional energy taken
up by electrons, $\epsilon_e$. $p$ here is the specral index of
the electron's energy distribution. Since typically $p \approx 2$
we have an almost  linear dependence of $L_x$ on $E_K$.  $L_x$
should be measured several hours after the burst before any jet
break takes place.

One can attempt to estimate $E_K$ from measurements of $L_x$.
However a simpler and more robust calculation will be to estimate
the width of the energy distribution $\sigma_{E_K}$ which in turn
is determined by the width of the x-ray luminosity distribution.
Remarkably, the  width of the x-ray luminosity distribution can be
estimated from the observed x-ray flux distribution. The x-ray
afterglow fluxes from GRBs have a power law dependence on $\nu$
and on the observed time $t$ (Piro, 2000): $f_\nu(t) \propto
\nu^{-\beta} t^{-\alpha}$ with $\alpha \sim 1.4$ and $\beta \sim
0.9$. The observed x-ray flux per unit frequency, $f_x$, is
related, therefore, to, $L_x$, the isotropic luminosity of the
source at redshift, z by:
\begin{equation}
L_x(t) = {4 \pi d_L^2 \over (1+z)} f_x(t)(1+z)^{\beta-\alpha}
\equiv f_{x}(t) Z(z) \ ,
\end{equation}
where $Z(z)$ is a weakly varying function of $z$. For bursts with
$0.5<z<4$ and with $\beta-\alpha \approx -0.5$ we find $\sigma_Z
\approx 0.31$ (for a cosmology with $\Omega_m=0.3$ and
$\Omega_\Lambda=0.7$). Here and thereafter we denote by
$\sigma_X$ the standard deviation of the $\log(X)$, unless noted
otherwise.

Piran et al (2001) use 21 BeppoSAX bursts (Piro, 2000) to
determine $\overline{\log(f_x)}$ and
$\sigma_{f_x}=0.43^{+.12}_{-.11}$ for the observed x-ray flux in
the 2--10 kev band at 11 hr after the  GRB   (For two of the 21
bursts there is only upper limit  of  $2 \times 10^{-13}{\rm
ergs/cm^2/sec}$ to the x-ray flux. This is consistent with a
predicted number of 3.5 burst with x-ray afterglows below this
limit.).

\begin{figure}
\plottwo{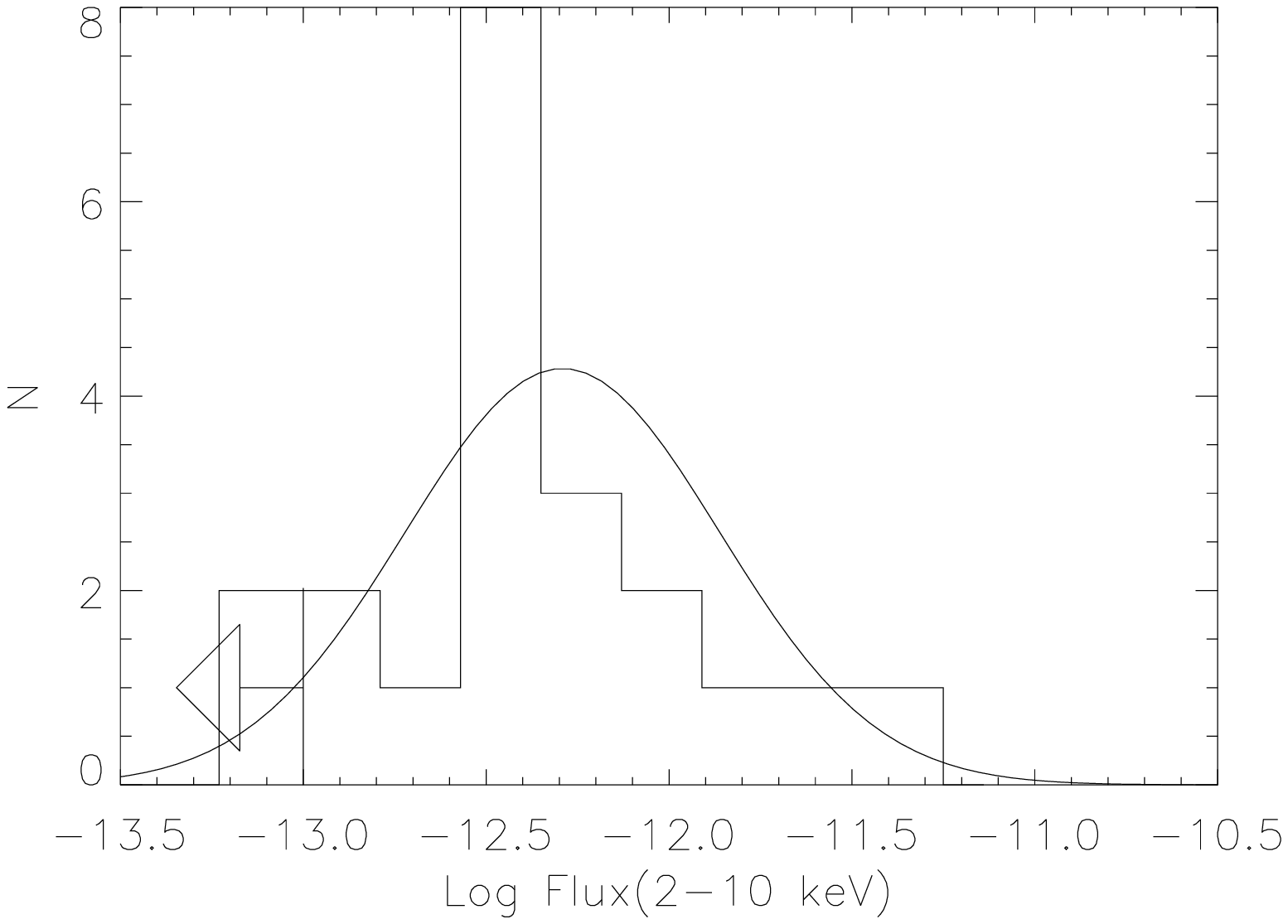}{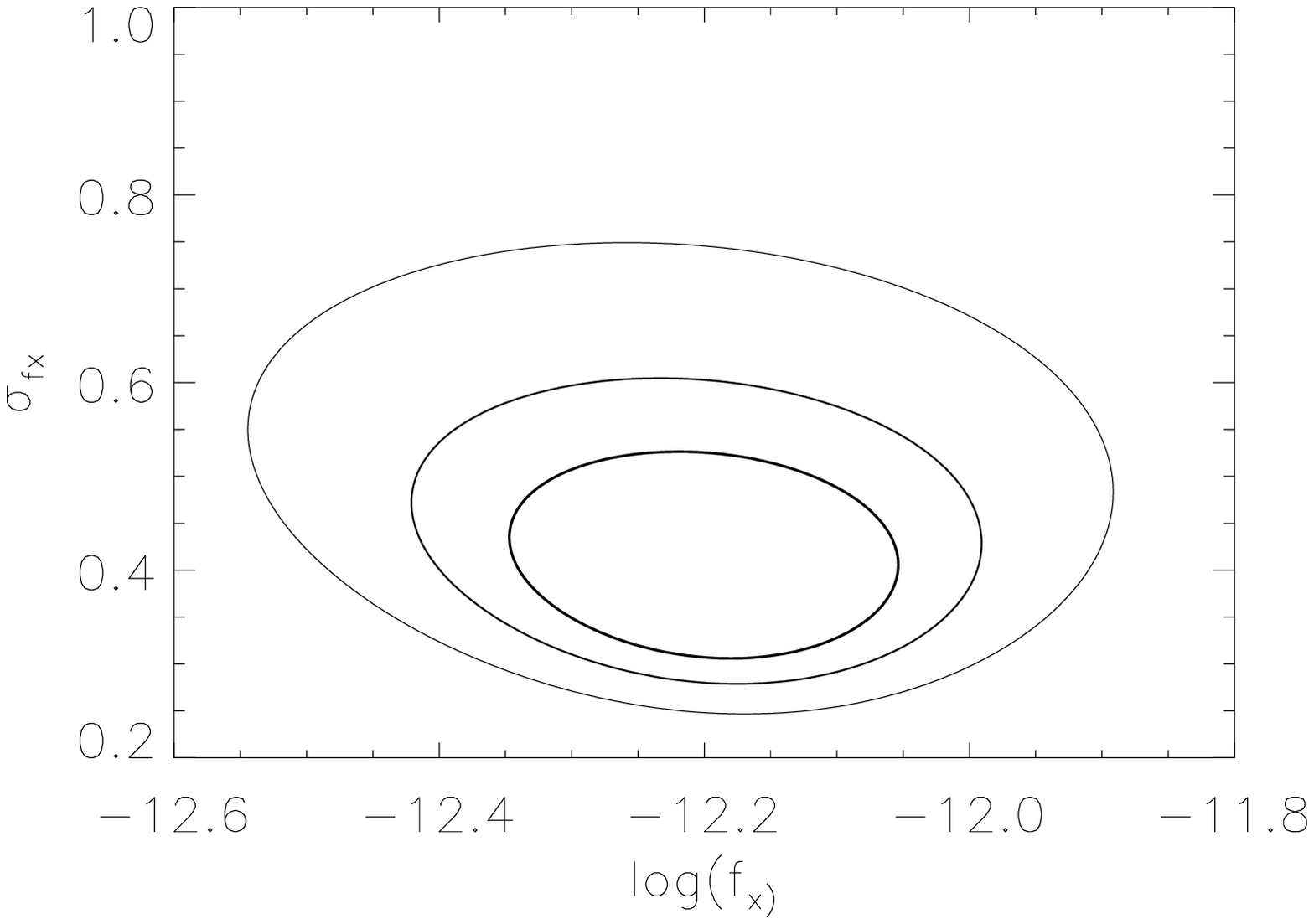} \caption{{\footnotesize From Piran
et al, (2001)  {\bf Left:} The distribution of X-ray fluxes (2-10
keV) at t=11 hours after the GRB in  21 afterglows observed by
BeppoSAX. The arrow marks two bursts for which there is only an
upper limit. {\bf Right:} Likelihood contour lines (corresponding
to 99\%, 90\% and 69\% confidence levels) in the
$\overline{\log(f_x)}$, $\sigma_{f_x}$ plane for the X-ray flux
distribution as inferred from 21 GRBs detected by BeppoSAX. The
maximal likelihood is at $\overline{\log(f_x)}=-12.2{\pm0.2}$ and
$\sigma_{f_x} =0.43^{+.12}_{-.11}$.}}
\end{figure}

If  there is no correlation between the microscopic variables
$\epsilon_e$, $\epsilon_B$, $p$ and $dE_K /d\Omega$ and if the
x-ray flux does not depend on the redshift we have:
\begin{equation}
\sigma^2_{E_K}+4 \sigma^2_\theta = \sigma^2_{dE_K/d\Omega} <
\sigma^2_{L_x}=\sigma^2_{f_{x}} + \sigma^2_Z\approx
\sigma^2_{f_{x}} = (0.43^{+.12}_{-.11})^2 \ .
\end{equation}
PK01 and F01 have estimated the jet opening angles  obtaining
$\sigma_\theta (PK01) \approx 0.31\pm 0.06$ and $\sigma_\theta
(F01) \approx 0.28\pm 0.05$ for 8 and 17 bursts respectively. If
these values are representative for the whole GRB population we
find a marginally viable solution within two $\sigma$ errors of
$\sigma_{E_K} < 0.2$ (for the PK01 result) and $\sigma_{E_K} <
0.27$ (for the F01 data); to get a viable solution we had to take
both the values of $\sigma_{L_x}$ one $\sigma$ above the mean and
the value of $\sigma_\theta$ one $\sigma$ below the mean. This
result suggests that there is a narrow energy distribution; the
FWHM of $E_K$ being less than a factor of 5.

We have argued before that $E_K$, discussed here, is a rather
good estimate to $E_{rel}$ the total energy emitted by the
``inner engine". The constancy of $E_K$ is another indication for
it being a good measure of $E_{rel}$. The constancy of $E_K$ is
also an indication that the assumptions that have lead to Eq. 2
are justified. Otherwise it would have been remarkable if
starting from different levels of initial energy and having
different amounts of energy losses the final kinetic energy of
the afterglow would converge to a constant value.

\section{Discussion}

There are two striking results: the narrowness of the $E_\gamma$
and $E_{K}$ distributions and the fact that  $\bar E_\gamma
\approx 3 \bar E_K$ (see Fig. 3). There seem to be an intrinsic
inconsistency between the two results. A narrow $E_\gamma$
distribution with $\bar E_\gamma
> \bar E_K$ implies (with no fine tuning) that  $E_{rel}$  is
narrowly distributed and  $E_\gamma \approx E_{rel}$ (rather than
$E_{rel} \approx E_K$). The fact that $E_K<E_\gamma$ is also
narrowly distributed implies that $\epsilon$, the conversion
efficiency of relativistic kinetic energy to $\gamma$-rays, is
close to unity and moreover, $\epsilon$ itself should be rather
narrowly distributed (between 70-80\%). This last conclusion
would be astonishing considering the dependence of $\epsilon$ on
the the distribution of energies and Lorentz factors of the
different shells. The narrow distribution of $E_\gamma$ also
implies that the emitting jets are rather homogeneous (otherwise
we would expect significant variations in $E_\gamma$ even for the
same burst when observed from slightly different direction). The
results are possible but they require a (currently) inexplicable
fine tuning.

There is no simple way out. It is of course possible that we have
been misled by small number statistics and we have to wait for a
larger data set in which we expect  to find $E_K \ge E_\gamma$ and
a larger spread in $E_\gamma$.

The current results would have been easier to understand if, for
some reason,  $E_K$  have been underestimated by a factor of 3.
In this case we would have $E_{rel} \approx E_\gamma \approx E_K$
with $\epsilon \approx 0.5$ and with a reasonable spread in
$\epsilon$. Note that at least in the PK01 results we already have
$\sigma_{E_\gamma} > \sigma_{E_K}$.

A second possibility is that in fact $E_K \ge \hat E_\gamma$
(where $\hat E_\gamma$ is the $\gamma$-ray energy averaged over
the whole GRB jet).  However, in our sample angular
inhomogeneities resulting in bright spots have increased,
according to the patchy shell model (Kumar \& Piran, 2000), the
observed $E_\gamma$ (in the afterglow sample) above the real
(average) $\hat E_\gamma$ value. This would results in
$\sigma_{E_\gamma}
> \sigma_{E_K}$, as observed indeed in the PK01 results. However,
the same angular inhomogeneities would result at times in a
decrease in the observed flux. We should observe also bursts with
$E_\gamma < E_K$. Such bursts are currently missing in the PK01
sample! Again more bursts may resolve this problem.

In any case  these results indicate that one way or another  GRB
"inner engines" are standard candles releasing a rather constant
energy. The wide distribution of directly and indirectly
determined $E_{\gamma,iso}$ results from the wide distribution of
beaming angles. The fact that GRB engines are "standard" engines
in terms of their energy output provide a very severe constraint
on the nature of these enigmatic explosions. For instance, in the
collapsar model for GRBs the central engine is composed of a
black hole (BH) and an accretion disk around it (Woosley 1993;
Paczynski, 1998; MacFadyen \& Woosley, 1999). This model has two
energy reservoirs which can be tapped to launch a relativistic
jet: the BH rotation energy and the gravitaional energy of the
disk. Our result of nearly constant energy in GRBs implies that
the mass accretion on to the BH plus the possible conversion of
rotational energy of the BH to kinetic energy of the jet does not
vary much from one burst to another in spite of the fact that
both the disk mass and the BH spin are expected to vary widely in
the collapse of massive stars.

I thank P. Kumar, E. Nakar A. Panaitescu and L. Piro for helpful
discussions. This research was partially supported by the
US-Israel BSF.


\begin{references}

\reference Beloborodov, A.~M.\ 2000, \apjl, 539, L25

\reference  Bloom, J.~S., Frail, D.~A., \& Sari, R.\ 2001, \aj,
121, 2879

\reference Burenin, R.A. et al., 1999, A \& A Supplement, 138, 443
\reference Cohen, E. and Piran, T., 1995, ApJ 444, L25

\reference Daigne, F.~\& Mochkovitch, R.\ 1998, \mnras, 296, 275

\reference Frail, D.A. et al. 2001, astro-ph/0102282, (F01)
\reference Freedman, D.L., and Waxman, E., 2001, ApJ 547, 922
\reference Giblin, T.W., et al. 1999, ApJ 524, L47 \reference
Granot J., Miller M.,  Piran T.,   Suen W-M \& Hughes P. A.,
2001,  proc of the II Rome Workshop on GRBs in the Afterglow Era
astro-ph/0103038


\reference Guetta, D., Spada, M., \& Waxman, E.\ 2001, \apj, 557,
399

\reference Jimenez R.,  Band D., \&  Piran, T. 2001, Ap. J. in
press, astro-ph/0103258
\reference Kobayashi, S., Piran, T., and
Sari, R., 1997, ApJ 490, 92
\reference Kobayashi, S.~\& Sari, R.\
2001, \apj, 551, 934

\reference Kulkarni et al. 1999, Nature 398, 389
\reference Kumar, P.\ 1999, \apjl, 523, L113

\reference Kumar, P., 2000, ApJ 538, L125
\reference Kumar, P.,\& Piran, T., 2000, ApJ 535, 152

\reference Lithwick, Y.~\& Sari, R.\ 2001, \apj, 555, 540


\reference MacFadyen, A. I. \& Woosley, S. E., 1999, ApJ, 524,262

\reference Meegan, C.A., et al., 1992, Nature, { 355}, 143

\reference R. Moderski, M. Sikora, T Bulik, 2000, Ap. J., {529},
151

\reference Paczynski, B., 1998, ApJ,494L, 45

\reference Panaitescu A., \&  M\'esz\'aros, P., 1999, Ap. J.,
{526},
  707

\reference Panaitescu, A., and Kumar, P., 2000, ApJ, 543, 66
\reference Panaitescu, A., and Kumar, P., 2001, ApJ, 560, 49
(PK01) \reference Piran, T. 1999, Physics Reports, 314, 575
 \reference
Piro, L., 2000, proc of X-Ray Astronomy 99: Stellar End points,
AGN and the Diffuse X-ray background, N. White ed., in press
astro-ph/0001436
 \reference Rhoads, J.E.,
1999, ApJ 525, 737

\reference Totani, T., 1997, Ap. J. Lett., {\bf 486}, 71
\reference Sahu, K., et al, 1997, Ap. J. Lett., {\bf 489}, L127
\reference Sari, R., Piran, T., and Halpern, J.P., 1999, ApJ 519,
L17

\reference Schmidt, M.\ 1999, \apjl, 523, L117

\reference Schmidt, M., 2001, ApJ 552, 36

\reference Tkachenko, A., et al. 2000, Astronomy \& Astrophysics,
358, L41

\reference Spada, M., Panaitescu, A., \& M{\' e}sz{\' a}ros, P.\
2000, \apj, 537, 824

\reference Wijers, R.A.M.J., et al, 1998, MNRAS, {\bf 294}, 13
\reference Woosley, S. E. 1993, ApJ, 405, 273

\end{references}
\end{document}